\documentclass{elsart}

\usepackage{amssymb}

\begin{document}

\begin{frontmatter}



\title{Quantum key distribution without alternative measurements and rotations\thanksref{sup}}
\thanks[sup]{Supported by the National Natural Science Foundation
of China, Grants No. 60373059; the National Laboratory for Modern
Communications Science Foundation of China, Grants No.
51436020103DZ4001; the National Research Foundation for the
Doctoral Program of Higher Education of China, Grants
No.20040013007; and the ISN Open Foundation.}


\author[BUPT,XD]{Fei Gao\corauthref{cor}},
\corauth[cor]{Corresponding author.} \ead{hzpe@sohu.com}
\author[BUPT]{Fenzhuo Guo},
\author[BUPT]{Qiaoyan Wen},
\author[CD]{Fuchen Zhu}

\address[BUPT]{School of Science, Beijing University of Posts and Telecommunications, Beijing, 100876, China}
\address[XD]{State Key Laboratory of Integrated Services Network, Xidian University, Xi'an, 710071, China}
\address[CD]{National Laboratory for Modern Communications, P.O.Box 810, Chengdu, 610041, China}

\begin{abstract}
A quantum key distribution protocol based on entanglement swapping
is proposed. Through choosing particles by twos from the sequence
and performing Bell measurements, two communicators can detect
eavesdropping and obtain the secure key. Because the two particles
measured together are selected out randomly, we need neither
alternative measurements nor rotations of the Bell states to
obtain security.
\end{abstract}
\end{frontmatter}

\section{Introduction}
\label{sec:level1} As a kind of important resource, entanglement
\cite {EPR} is widely used in the research of quantum information,
including quantum communication, quantum cryptography and quantum
computation. Entanglement swapping \cite{ZZHE}, abbreviated by ES,
is a nice property of entanglement. That is, by appropriate Bell
measurements, entanglement can be swapped between different
particles. For example, consider two pairs of particles in the
state of $|\Phi^+\rangle$, equivalently,
$|\Phi^+\rangle_{12}=|\Phi^+\rangle_{34}=1/\sqrt{2}(|00\rangle+|11\rangle)$,
where the subscripts denote different particles. If we make a Bell
measurement on 1 and 3, they will be entangled to one of the Bell
states. Simultaneously, 2 and 4 will be also projected onto a
corresponding Bell state. We can find the possible results through
the following process:
\begin{eqnarray}
|\Phi^+\rangle_{12}\otimes|\Phi^+\rangle_{34}&=&\frac{1}{2}(|00\rangle+|11\rangle)_{12}
\otimes(|00\rangle+|11\rangle)_{34}\nonumber\\
&=&\frac{1}{2}(|0000\rangle+|0101\rangle +|1010\rangle
+|1111\rangle)_{1324}\nonumber\\
&=&\frac{1}{2}(|\Phi^+\Phi^+\rangle+|\Phi^-\Phi^-\rangle
+|\Psi^+\Psi^+\rangle+|\Psi^-\Psi^-\rangle)_{1324}
\end{eqnarray}
It can be seen that there are four possible results:
$|\Phi^+\rangle_{13}|\Phi^+\rangle_{24}$,
$|\Phi^-\rangle_{13}|\Phi^-\rangle_{24}$,
$|\Psi^+\rangle_{13}|\Psi^+\rangle_{24}$ and
$|\Psi^-\rangle_{13}|\Psi^-\rangle_{24}$. Furthermore, these
results appear with equal probability, that is, $1/4$. For further
discussion about ES, please see Refs.\cite {ZHWZ,BVK,KBB,PBWZ}.

Quantum cryptography is the combination of quantum mechanics and
cryptography. It employs fundamental theories in quantum mechanics
to obtain unconditional security. Quantum key distribution (QKD)
is an important research direction in quantum cryptography.
Bennett and Brassard came up with the first QKD protocol (BB84
protocol) in 1984 \cite{BB84}. Afterwards, many protocols were
presented \cite
{E91,B92,BW92,GV95,HIGM,KI97,B98,LL02,PBTB,XLG,LCA,CHL,SJL,J1,J2}.
Recently, several QKD schemes based on ES were proposed
\cite{C2000,ZLG,C2001,D2004,CQ-PH,LLKO,ZYCP,LSZW}. In
Refs.\cite{C2000,ZLG,C2001} the author introduced a protocol
without alternative measurements. It was simplified \cite{D2004}
and generalized \cite{CQ-PH} before long, and its security was
proved in Ref.\cite{LLKO}. Besides, by ES, doubly entangled photon
pairs \cite{ZYCP} and previously shared Bell states \cite{LSZW}
can be used to distribute secure key.

In this Letter we propose a QKD protocol based on ES, which needs
neither alternative measurements \cite{ZYCP} nor rotations of the
Bell states \cite{C2001,D2004,CQ-PH}. The security against the
attack discussed in Ref.\cite{ZLG} is assured by a special
technique, that is, random grouping (RG). See Sec.2 for the
details of this protocol. The security against general individual
attack is analyzed in Sec.3 and a conclusion is given in Sec.4.

\section{The QKD protocol}
The particular process of this scheme is as follows:

1. Prepare the particles. Alice generates a sequence of EPR pairs
in the state
$|\Phi^+\rangle_{AB}=1/\sqrt{2}(|00\rangle+|11\rangle)$. For each
pair, Alice stores one particle and sends the other to Bob.

2. Detect eavesdropping.

(1) Having received all the particles from Alice, Bob randomly
selects a set of particles out and makes Bell measurements on them
by twos.

(2) Bob tells Alice the sequence numbers and measurement results
of the pairs he measured.

(3) According to the sequence numbers, Alice performs Bell
measurements on the corresponding pairs, and compares her results
with Bob's. For example, consider one of the pairs Bob measured,
in which the sequence numbers of the two particles are $m$ and
$n$, respectively. Then Alice measures her $m$-th and $n$-th
particles in Bell basis, and compares the two outcomes. As
discussed in Sec.1, if these particles were not eavesdropped,
Alice and Bob should obtain the same results. With this knowledge,
Alice can determine, through the error rate, whether there is any
eavesdropping. If there are no eavesdroppers in the channel, Alice
and Bob proceed with the next step.

3. Obtain the key. Bob makes Bell measurements on his left
particles by twos. It should be emphasized that each pair he
measures is selected out randomly. Bob records the sequence
numbers of all these pairs and sends the record to Alice. Alice
then measures her corresponding particles in Bell basis. As
discussed in the above paragraphs, their measurement results would
be identical. Subsequently, Alice and Bob can obtain the raw key
from these results. For example, $|\Phi^+\rangle$,
$|\Phi^-\rangle$, $|\Psi^+\rangle$ and $|\Psi^-\rangle$ are
encoded into $00$, $01$, $10$ and $11$, respectively. After error
correction and privacy amplification \cite{GRTZ}, the raw key will
be processed into ideal secret key.

Thus the whole QKD protocol is finished. By this process, Alice
and Bob can obtain secure key. In this scheme, we use
$|\Phi^+\rangle$ as the initial state. In practice, any other Bell
state is competent and the communicators can even utilize various
states for different pairs. It should be emphasized that, however,
the various initial states cannot improve the efficiency of QKD
(the alleged ``high efficiency'' in Ref.\cite{LSZW} is a mistake
\cite{QWZ}). In fact, our protocol works in a deterministic manner
and then has full efficiency in the sense that one
qubit-transmission brings one key bit. That is, except for the
detection particles, the users can obtain 1 bit (raw) key per
qubit-transmission in our protocol, which is higher than the BB84
protocol (0.5 bit).

To compare the efficiency of our protocol with that of others
deeply, we can employ Cabello's definition of QKD efficiency
\cite{CHL}. Let us give a simple example to implement the above
protocol and then calculate its efficiency. Suppose Alice and Bob
deal with four EPR pairs (denoted as pairs 1,2,3,4, respectively)
in one step. More specifically, Alice sends four particles (each
from one of the four EPR pairs) to Bob and announces a classical
(random) bit (0 or 1) after Bob received this group of particles.
If the classical bit is 0, they perform ES on the pairs 1,3 and
2,4 to obtain the key. Otherwise they perform ES on the pairs 1,4
and 2,3. In this example, Alice and Bob get four key bits by
transmitting four qubit and one cbit (classical bit). Obviously,
the efficiency equals to 0.8, which is relatively higher (For
instance, the efficiency of the famous protocols in
Ref.\cite{B92}, \cite{BB84}, \cite{GV95}, \cite{E91}, \cite{KI97},
\cite{C2000} is $<0.25$, $0.25$, $\leq0.33$, $0.5$, $0.5$, $0.67$,
respectively. See Table I in Ref.\cite{CHL} for details).
\section{Security}
The above scheme can be regarded as secure because the key
distributed can not be eavesdropped imperceptively. There are two
general eavesdropping strategies for Eve. One is called
``intercept and resend'', that is, Eve intercepts the legal
particles and replaces them by her counterfeit ones. For example,
Eve generates the same EPR pairs and sends one particle from each
pair to Bob, thus she can judge Bob's measurement results as Alice
does in step 3. But in this case there are no correlations between
Alice's particles and the counterfeit ones. Alice and Bob will get
random measurement results when they detect eavesdropping in step
2. Suppose both Alice and Bob use $s$ pairs to detect
eavesdropping, the probability with which they obtain the same
results is only $(1/4)^s$ . That is, Eve will be detected with
high probability when $s$ is big enough. The second strategy for
Eve is to entangle an ancilla with the two-particle state that
Alice and Bob are using. At some later time she can measure the
ancilla to gain information about the measurement results of Bob.
This kind of attack seems to be stronger than the first strategy.
However, it is invalid to our protocol as we prove below.

Because each particle transmitted in the channel is in a maximally
mixed state, there are no differences among all these particles
for Eve. Furthermore, Eve does not know which two particles Bob
will put together to make a Bell measurement. As a result, what
she can do is to make the same operation on each particle. Let
$|\varphi\rangle_{ABE}$ denote the state of the composite system
including one certain EPR pair and the corresponding ancilla,
where the subscripts $A$, $B$ and $E$ express the particles
belonging to Alice, Bob and Eve, respectively. Note that each
ancilla's dimension is not limited here, and Eve is permitted to
build all devices allowed by the laws of quantum mechanics. What
we want to show is that $|\varphi\rangle_{ABE}$ must be a product
of a two-particle state and the ancilla if the eavesdropping
introduces no errors into the QKD procedure, which implies that
Eve will gain no information about the key by observing the
ancilla. Conversely, if gaining information about the key, Eve
will invariably introduce errors.

Without loss of generality, suppose the Schmidt decomposition
\cite{QCQI} of $|\varphi\rangle_{ABE}$  is in the form

\begin{eqnarray}
|\varphi\rangle_{ABE}=a_1|\psi_1\rangle_{AB}|\phi_1\rangle_E
                     +a_2|\psi_2\rangle_{AB}|\phi_2\rangle_E\nonumber\\
                     +a_3|\psi_3\rangle_{AB}|\phi_3\rangle_E
                     +a_4|\psi_4\rangle_{AB}|\phi_4\rangle_E
\end{eqnarray}
where $|\psi_i\rangle$ and $|\phi_j\rangle$  are two sets of
orthonomal states, $a_k$ are non-negative real numbers
($i,j,k=1,2,3,4$ ).

Because $|\psi_i\rangle$ are two-particle (four-dimensional)
states, they can be written as the linear combinations of
$|00\rangle$, $|01\rangle$, $|10\rangle$ and $|11\rangle$. Let
\begin{eqnarray}
|\psi_1\rangle=b_{11}|00\rangle+b_{12}|01\rangle+b_{13}|10\rangle+b_{14}|11\rangle\nonumber\\
|\psi_2\rangle=b_{21}|00\rangle+b_{22}|01\rangle+b_{23}|10\rangle+b_{24}|11\rangle\nonumber\\
|\psi_3\rangle=b_{31}|00\rangle+b_{32}|01\rangle+b_{33}|10\rangle+b_{34}|11\rangle\nonumber\\
|\psi_4\rangle=b_{41}|00\rangle+b_{42}|01\rangle+b_{43}|10\rangle+b_{44}|11\rangle
\end{eqnarray}
where $b_{pq}$ ($p,q=1,2,3,4$) are complex numbers. Then
$|\varphi\rangle_{ABE}$ can be written, thanks to Eqs.(2) and (3),
as
\begin{eqnarray}
|\varphi\rangle_{ABE}=|00\rangle_{AB}\otimes(a_1b_{11}|\phi_1\rangle+a_2b_{21}|\phi_2\rangle+a_3b_{31}|\phi_3\rangle+a_4b_{41}|\phi_4\rangle)_E\nonumber\\
                     +|01\rangle_{AB}\otimes(a_1b_{12}|\phi_1\rangle+a_2b_{22}|\phi_2\rangle+a_3b_{32}|\phi_3\rangle+a_4b_{42}|\phi_4\rangle)_E\nonumber\\
                     +|10\rangle_{AB}\otimes(a_1b_{13}|\phi_1\rangle+a_2b_{23}|\phi_2\rangle+a_3b_{33}|\phi_3\rangle+a_4b_{43}|\phi_4\rangle)_E\nonumber\\
                     +|11\rangle_{AB}\otimes(a_1b_{14}|\phi_1\rangle+a_2b_{24}|\phi_2\rangle+a_3b_{34}|\phi_3\rangle+a_4b_{44}|\phi_4\rangle)_E
\end{eqnarray}

For convenience, we define four vectors (not quantum states) as
follows:
\begin{equation}
v_l=(a_1b_{1l},a_2b_{2l},a_3b_{3l},a_4b_{4l}) \quad l=1,2,3,4
\end{equation}
Consider any two sets of particles on which Alice and Bob will do
ES, the state of the system is
$|\varphi\rangle_{ABE}\otimes|\varphi\rangle_{ABE}$. According to
the properties of ES, we can calculate the probability with which
each possible measurement-results-pair is obtained after Alice and
Bob measured their particles in Bell basis. For example, observe
the event that Alice gets $|\Phi^+\rangle$ and Bob gets
$|\Psi^+\rangle$ , which corresponds to the following item in the
expansion:
\begin{equation}
\frac{1}{2}|\Phi^+\rangle_A|\Psi^+\rangle_B\otimes\left[\sum_{r,s=1}^4
(a_rb_{r1}a_sb_{s2}+a_rb_{r2}a_sb_{s1}+a_rb_{r3}a_sb_{s4}+a_rb_{r4}a_sb_{s3})
|\phi_r\phi_s\rangle_E\right]
\end{equation}
Therefore, this event occurs with the probability
\begin{equation}
P(\Phi_A^+\Psi_B^+)=\frac{1}{4}\sum_{r,s=1}^4|a_rb_{r1}a_sb_{s2}+a_rb_{r2}a_sb_{s1}+a_rb_{r3}a_sb_{s4}+a_rb_{r4}a_sb_{s3}|^2
\end{equation}
However, this event should not occur. In fact, if Eve wants to
escape from the detection of Alice and Bob, any results-pair other
than $\Phi^+\Phi^+$, $\Phi^-\Phi^-$, $\Psi^+\Psi^+$ and
$\Psi^-\Psi^-$ should not appear. Let $P(\Phi_A^+\Psi_B^+)=0$, we
then have, from Eqs.(7) and (5),
\begin{equation}
v_1^Tv_2+v_2^Tv_1+v_3^Tv_4+v_4^Tv_3=0
\end{equation}
in which $v_l^T$ is the transpose of $v_l$.

Similarly, let the probabilities of $\Phi_A^+\Psi_B^-$,
$\Phi_A^-\Psi_B^+$ and $\Phi_A^-\Psi_B^-$ equal to 0, we get
\begin{equation}
v_1^Tv_2-v_2^Tv_1+v_3^Tv_4-v_4^Tv_3=0
\end{equation}
\begin{equation}
v_1^Tv_2+v_2^Tv_1-v_3^Tv_4-v_4^Tv_3=0
\end{equation}
\begin{equation}
v_1^Tv_2-v_2^Tv_1-v_3^Tv_4+v_4^Tv_3=0
\end{equation}
From Eqs.(8)-(11), we can obtain
\begin{equation}
v_1^Tv_2=v_2^Tv_1=v_3^Tv_4=v_4^Tv_3=0
\end{equation}
That is,
\begin{eqnarray}
\left\{\begin{array}{c}
v_1=0 \quad or \quad v_2=0\\
v_3=0 \quad or \quad v_4=0
\end{array}\right.
\end{eqnarray}
For the same reason, we can obtain the following results:\\
(1) Let the probabilities of $\Psi_A^+\Phi_B^+$,
$\Psi_A^+\Phi_B^-$, $\Psi_A^-\Phi_B^+$ and $\Psi_A^-\Phi_B^-$
equal to 0, we can get
\begin{eqnarray}
\left\{\begin{array}{c}
v_1=0 \quad or \quad v_3=0\\
v_2=0 \quad or \quad v_4=0
\end{array}\right.
\end{eqnarray}
(2) Let the probabilities of $\Phi_A^+\Phi_B^-$  and
$\Phi_A^-\Phi_B^+$ equal to 0, we then have
\begin{equation}
v_1^Tv_1-v_2^Tv_2+v_3^Tv_3-v_4^Tv_4=0
\end{equation}
\begin{equation}
v_1^Tv_1+v_2^Tv_2-v_3^Tv_3-v_4^Tv_4=0
\end{equation}
And then
\begin{eqnarray}
\left\{\begin{array}{c}
v_1=\pm v_4\\
v_2=\pm v_3
\end{array}\right.
\end{eqnarray}
(3) Let the probabilities of $\Psi_A^+\Psi_B^-$ and
$\Psi_A^-\Psi_B^+$ equal to 0, we can get the same conclusion as
Eq.(17).

Finally, we can obtain three results from Eqs.(13), (14) and (17):
\begin{description}
\item[\quad 1.] $v_1=v_2=v_3=v_4=0$ ; \item[\quad 2.]$v_1=v_4=0$
and $v_2=\pm v_3$; \item[\quad 3.]$v_2=v_3=0$ and $v_1=\pm v_4$
\end{description}
That is, each of these results makes Eve succeed in escaping from
the detection of Alice and Bob. Now we observe what the state
$|\varphi\rangle_{ABE}$ is by putting these results into Eq.(4).
If the first result holds, we have $|\varphi\rangle_{ABE}=0$,
which is meaningless for our analysis. Consider the condition
where the second result holds, $|\varphi\rangle_{ABE}$ can be
written as:
\begin{eqnarray}
|\varphi\rangle_{ABE}&=&(|01\rangle\pm|10\rangle)_{AB}\otimes(a_1b_{12}|\phi_1\rangle\nonumber\\
&+&a_2b_{22}|\phi_2\rangle+a_3b_{32}|\phi_3\rangle+a_4b_{42}|\phi_4\rangle)_E
\end{eqnarray}
It can be seen that $|\varphi\rangle_{ABE}$ is a product of a
two-particle state and the ancilla. That is, there is no
entanglement between Eve's ancilla and the legal particles, and
Eve can obtain no information about the key. Similarly, we can
draw the same conclusion when the third result holds.

From another point of view, we can derive an effective relation
between the errors introduced in the key and the information
gained by Eve as in Ref.\cite{CL}. Consider any two EPR pairs on
which Alice and Bob will perform ES, for example, $\Phi^+_{12}$
and $\Phi^+_{34}$, where particles 1, 3 and 2, 4 belong to Alice
and Bob respectively. As we know, when Alice and Bob make Bell
measurements on these particles, the marginal statistics of the
measurement results are independent of the measurement order.
Suppose Alice makes her measurement before Bob, the state of 2, 4
will thus be projected onto one of the Bell states $|\xi\rangle$.
Because of Eve's intervention, these two particles will be
entangled into Eve's ancilla and it follows that the state
$|\xi\rangle$ becomes a mixed state $\rho$. The information Bob
can gain from $\rho$ is bounded by the Holevo quantity
$\chi(\rho)$ \cite{QCQI}. Let $I_{Eve}$ denote the information Eve
can obtain, then $I_{Eve}\leq\chi(\rho)$. (Obviously, Eve can not
gain more information about Bob's measurement result than Bob.)
From
\begin{equation}
\chi(\rho)=S(\rho)-\sum_ip_iS(\rho_i)
\end{equation}
we know $S(\rho)$ is the upper bound of $\chi(\rho)$. ``High
fidelity implies low entropy" \cite{CL}. Suppose
\begin{equation}
F(|\xi\rangle,\rho)^2=\langle\xi|\rho|\xi\rangle=1-\gamma
\end{equation}
where $F(|\xi\rangle,\rho)$ is the fidelity \cite{Fuchs} of the
states $|\xi\rangle$ and $\rho$, $0\leq\gamma\leq1$. Therefore,
the entropy of $\rho$ is bounded above by the entropy of a
diagonal density matrix $\rho_{max}$ with diagonal entries
$1-\gamma$, $\gamma/3$, $\gamma/3$, $\gamma/3$. The entropy of
$\rho_{max}$ is
\begin{equation}
S(\rho_{max})=-(1-\gamma)\log_2(1-\gamma)-\gamma\log_2\frac{\gamma}{3}
\end{equation}
Then we have
\begin{equation}
I_{Eve}\leq-(1-\gamma)\log_2(1-\gamma)-\gamma\log_2\frac{\gamma}{3}
\end{equation}

Let us discuss the connection between the fidelity
$F(|\xi\rangle,\rho)$ and the detection probability $d$. When
Alice and Bob detect eavesdropping, only $|\xi\rangle$ is the
correct result, whereas any other Bell state will be regarded as
an error. Since $F(|\xi\rangle,\rho)^2=1-\gamma$, the detection
probability $d=\gamma$. From Eq.(22), we get
\begin{equation}
I_{Eve}\leq-(1-d)\log_2(1-d)-d\log_2\frac{d}{3}
\end{equation}

It can be seen from this relation that when $d=0$, i.e., Eve
introduces no error to the key, she will obtain no information,
which is in agreement with the above result. When $\gamma>0$,
i.e., Eve can gain some of Bob's information, but she has to face
a nonzero risk $d=\gamma$ of being detected. When $\gamma=3/4$, we
have $S(\rho_{max})=2$, which implies that Eve has the chance to
eavesdrop on all of Bob's information. In this case, however, the
detection probability is no less than $3/4$ per ES for
eavesdropping detection. For example, when Eve intercepts all the
particles and resends new particles from her own EPR pairs, she
will get all of the information about Bob's key while introduce
$3/4$ error rate per ES.

To sum up, our protocol can resist the eavesdropping with ancilla.

\section{Conclusion}
We have presented a full-efficiency QKD protocol based on ES. The
security against the attack discussed in Ref.\cite{ZLG} is assured
by the technique of RG instead of requiring alternative
measurements \cite{ZYCP} or rotations of the Bell states
\cite{C2001,D2004,CQ-PH}. Furthermore, this technique brings us
another advantage. That is, it is unnecessary to randomize the
initial Bell states as in Refs.\cite{C2000,C2001}, which leads to
less Bell measurements in our protocol. For instance, to
distribute two key bits, Alice and Bob make two Bell measurements
in our protocol, while in Refs.\cite{C2000,C2001} they must make
three.

On the other hand, we have to confess that our protocol has a
disadvantage, i.e., it uses a sequence of entangled states instead
of a single quantum system \cite{C2001,D2004,CQ-PH} to generate
the key. Nevertheless, it is not a fatal problem. Many QKD
protocols work in this model, for example, the famous E91 protocol
\cite{E91}. Furthermore, each pair of particles is still in one of
the Bell states and can be reused in other applications after QKD.

In practical implementations, our scheme needs complete Bell
states analysis. Though Bell measurement has not been generally
accomplished \cite{LCS}, it was experimentally realized based on
some certain techniques \cite{K2003,WPM,BKN}. Furthermore, the
realizations of entanglement swapping has been proposed
\cite{PBWZ,RMH}. Therefore, our scheme is within the reach of
current technology.

\end{document}